\documentclass[epj
]{svjour}
\usepackage{graphicx}
\begin{document}
\titlerunning{Pattern Formation in Games with Success-Driven Motion}
\title{Pattern Formation, Social Forces, and Diffusion Instability in Games with Success-Driven Motion}
\author{Dirk Helbing
}                     
%
%
\institute{ETH Zurich, UNO D11, Universit\"atstr. 41, 8092 Zurich, Switzerland}
\date{Received: date / Revised version: date}
%
\abstract{A local agglomeration of cooperators can support the survival or spreading of cooperation, even when cooperation is predicted to die out according to the replicator equation, which is often used in evolutionary game theory to study the spreading and disappearance of strategies. In this paper, it is shown that success-driven motion can trigger such local agglomeration and may, therefore, be used to supplement other mechanisms supporting cooperation, like reputation or punishment. Success-driven motion is formulated here as a function of the game-theoretical payoffs. It can change the outcome and dynamics of spatial games dramatically, in particular as it causes attractive or repulsive interaction forces. These forces act when the spatial distributions of strategies are inhomogeneous. However, even when starting with homogeneous initial conditions, small perturbations can trigger large inhomogeneities by a pattern-formation instability, when certain conditions are fulfilled. Here, these instability conditions are studied for the prisoner's dilemma and the snowdrift game. 
Furthermore, it is demonstrated that asymmetrical diffusion can drive social, economic, and biological systems into the unstable regime, if these would be stable without diffusion.
\PACS{
{02.50.Le}{Decision theory and game theory} \and
{87.23.Ge}{Dynamics of social systems} \and
{82.40.Ck}{Pattern formation in reactions with diffusion, flow and heat transfer} \and
{87.23.Cc}{Population dynamics and ecological pattern formation} 
} 
} 
\maketitle
\section{Introduction}

Game theory is a well-established theory of individual strategic interactions with applications in sociology, economics, and biology  \cite{Neumann,Axelrod,Rapoport,Gintis,Binmore,Nowak}, and with many publications even in physics (see Ref. \cite{Network2} for an overview). It distinguishes different behaviors, so-called strategies $i$, and expresses the interactions of individuals in terms of payoffs $P_{ij}$. The value $P_{ij}$ quantifies the result of an interaction between strategies $i$ and $j$ for the individual pursuing strategy $i$. The more favorable the outcome of the interaction, the higher is the payoff $P_{ij}$.
\par
There are many different games, depending on the structure of the payoffs,  the social interaction network, the number of interaction partners, the frequency of interaction, and so on \cite{Gintis,Binmore}. Theoretical predictions for the selection of strategies mostly assume a rational choice approach, i.e. a payoff maximization by the individuals, although experimental studies \cite{Kagel,Camerer,Henrich} support conditional cooperativity \cite{Conditional} and show that moral sentiments \cite{Unselfish} can support cooperation. Some models also take into account learning (see, e.g. \cite{Flache} and references therein), where it is common to assume that more successful behaviors are imitated (copied). Based on a suitable specification of the imitation rules, it can be shown \cite{imit1,imit2,Kluwer} that the resulting dynamics can be described by game-dynamical equations \cite{gamedyn1,gamedyn2}, which agree with replicator equations for the fitness-dependent reproduction of individuals in biology \cite{Eigen,Fisher,Schuster}.
\par
Another field where the quantification of human behavior in terms of mathematical models has been extremely successful concerns the dynamics of pedestrians \cite{molnar}, crowds \cite{panic}, and traffic \cite{tilch}. The related studies have led to fundamental insights into observed self-organization phenomena such as stop-and-go waves \cite{tilch} or lanes of uniform walking direction \cite{molnar}. In the meantime, there are many empirical \cite{crowdturb} and experimental results \cite{TranSci,Hoogendoorn}, which made it possible to come up with well calibrated models of human motion \cite{ACS3,Yu}.
\par
Therefore, it would be interesting to know what happens if game theoretical models are combined with models of driven motion. Would we also observe self-organization phenonomena in space and time? 
This is the main question addressed in this paper. Under keywords such as ``assortment'' and ``network reciprocity'', it has been discussed that the clustering of cooperators can amplify cooperation, in particular in the prisoner's dilemma \cite{Pepper,cluster1,cluster2,cluster3}. Therefore, the pattern formation instability is of prime importance to understand the emergence of cooperation between individuals. In Sec. \ref{results}, we will study the instability conditions for the prisoner's dilemma and the snowdrift game. Moreover, we will see that games with success-driven motion and asymmetrical diffusion may show pattern formation, where a homogeneous distribution of strategies would be stable {\it without} the presence of diffusion. It is quite surprising that sources of noise like diffusion can support the self-organization in systems, which can be described by game-dynamical equations with success-driven motion. This includes social, economic, and biological systems.
\par
We now proceed as follows: In Sec. \ref{PrisDil}, we introduce the game-dynamical replicator equation for the prisoner's dilemma (PD) and the snow-drift game (SD). In particularly, we discuss the stationary solutions and their stability, with the conclusion that cooperation is expected to disappear in the prisoner's dilemma. In Sec. \ref{withSpace}, we extend the game-dynamical equation by the consideration of spatial interactions, success-driven motion, and diffusion. Section \ref{COMPA} compares the resulting equations with reaction-diffusion-advection equations and discusses the similarities and differences with Turing instabilities and differential flow-induced chemical instabilities (DIFICI). Afterwards, Sec. \ref{results} analyzes the pattern formation instability for the prisoner's dilemma and the snowdrift game with its interesting implications, while details of the instability analysis are provided in Appendix \ref{BBB}. Finally, Sec. \ref{forces} studies the driving forces of the dynamics in cases of large deviations from stationary and homogeneous strategy distributions, before Sec. \ref{summa} summarizes the paper and presents an outlook.

\section{The Prisoner's Dilemma without Spatial Interactions}\label{PrisDil}

In order to grasp the major impact of success-driven motion and diffusion on the dynamics of games (see Sec. \ref{withSpace}), it is useful to investigate first the game-dynamical equations without spatial interactions. For this, we represent the proportion of individuals using a strategy $i$ at time $t$ by $p_i(t)$. While the discussion can be extended to any number of strategies, we will focus on the case of two strategies only for the sake of analytical tractability. Here, $i=1$ shall correspond to the cooperative strategy, $i=2$ to defection (cheating or free-riding). According to the definition of probabilities, we have $0\le p_i(t) \le 1$ for $i\in\{1,2\}$ and the normalization condition
\begin{equation}
p_1(t) + p_2(t) = 1 \, .
\label{normalize}
\end{equation}
Let $P_{ij}$ be the payoff, if strategy $i$ meets strategy $j$. Then, the expected payoff for someone applying strategy $i$ is
\begin{equation}
E_i(t) = \sum_{j=1}^2 P_{ij} p_j(t) \, ,
\label{compeq}
\end{equation}
as $p_j(t)$ represents the proportion of strategy $j$, with which the payoffs $P_{ij}$ must be weighted. The average payoff in the population of individuals is
\begin{equation}
\overline{E}(t) = \sum_{l=1}^2 E_l(t) p_l(t) = \sum_{l=1}^2 \sum_{j=1}^2 p_l(t) P_{lj} p_j(t) \, .
\end{equation}
In the game-dynamical equations, the temporal increase $dp_i(t)/dt$ of the proportion of individuals using strategy $i$ is proportional to the number of individuals pursuing strategy $i$ who may imitate, i.e. basically to $p_i(t)$. The proportionality factor, i.e. the growth rate $\lambda(i,t)$, is given by the difference between the expected payoff $E_i(t)$ and the {\it average} payoff $\overline{E}(t)$:
\begin{eqnarray}
 \frac{dp_i(t)}{dt} &=& \lambda(i,t) p_i(t) = \big[E_i(t) - \overline{E}(t)\big] p_i(t) \nonumber \\
 &=&  \bigg( \sum_{j=1}^2 P_{ij} p_j(t) - \sum_{l=1}^2\sum_{j=1}^2  p_l(t) P_{lj} p_j(t) \bigg) p_i(t) \, . \quad 
\label{gamedyn}
\end{eqnarray}
The equations (\ref{gamedyn}) are known as replicator equations. They were originally developed in evolutionary biology to describe the spreading of ``fitter'' individuals through their higher reproductive success \cite{Eigen,Fisher,Schuster}. However, the replicator equations were also used in game theory, where they are called ``game-dynamical equations''  \cite{gamedyn1,gamedyn2}. For a long time, it was not clear whether or why these equations could be applied to the frequency $p_i(t)$ of behavioral strategies, but it has been shown that the equations can be derived from Boltzmann-like equations for imitative pair interactions of individuals, if ``proportional imitation'' or similar imitation rules are assumed \cite{imit1,imit2,Kluwer}.
\par
Note that one may add a mutation term to the right-hand side of the game-dynamical equations (\ref{gamedyn}). This term could, for example, be specified as
\begin{eqnarray}
 W_2 p_2(t) - W_1 p_1(t) 
 &=& rq [1-p_1(t)] - r(1-q) p_1(t) \nonumber \\
 &=& r [ q - p_1(t) ]  
 \label{mutant}
\end{eqnarray}
for strategy $i=1$, and by the negative expression of this for $i=2$. Here,
$r$ is the overall mutation rate, $W_2=rq$ the mutation rate towards cooperation, and $W_1=r(1-q)$ the mutation rate towards defection \cite{Kluwer}. This implementation reflects spontaneous, random strategy choices due to erroneous or exploration behavior and modifies the stationary solutions.
\par
It can be shown that 
\begin{equation}
\sum_{i=1}^2 \frac{dp_i(t)}{dt} = 0 \, ,
\end{equation}
so that the normalization condition (\ref{normalize}) is fulfilled at all times $t$, if it is fulfilled at $t=0$. Moreover, the equation $dp_i(t)/dt = \lambda p_i(t)$ implies $p_i(t)\ge 0$ at all times $t$, if $p_i(0)\ge 0$ for all strategies $i$.
\par
We may now insert the payoffs of the prisoner's dilemma, i.e. 
\begin{eqnarray}
P_{11} &=& R \mbox{ (``reward''),} \nonumber \\
P_{12} &=& S \mbox{ (``sucker's payoff''),} \nonumber \\
P_{21} &=& T \mbox{ (``temptation''), and}\nonumber \\
P_{22} &=& P \mbox{ (``payoff'')} \label{PAYOFF}
\end{eqnarray}
with the assumed payoff relationships
\begin{equation}
T > R > P > S  \, .
\label{condi}
\end{equation}
Additionally, one often requires 
\begin{equation}
2R > S + T \, .\label{xcv}
\end{equation}
The ``reward'' $R$ is the payoff for mutual cooperation and the ``punishment'' $P$ the payoff for mutual defection, while $T$  is the ``temptation'' of unilateral defection, and a 
cheated cooperator receives the sucker's payoff $S$. While we have $P>S$ in the prisoner's dilemma, the snowdrift game (also known as chicken or hawk-dove game) is characterized by $S>P$, i.e. it is defined by
\begin{equation}
T > R > S > P \, . 
\label{snow}
\end{equation}
Both games are characterized by a temptation to defect ($T>R$), while the prisoner's dilemma has the additional challenge that there is a high risk to cooperate ($S<P$). This difference has a large influence on the resulting level $p_1(t)$ of cooperation: Inserting the above payoffs into Eq. (\ref{gamedyn}), one eventually obtains the game-dynamical equation
\begin{equation}
\frac{dp_1(t)}{dt} = [1-p_1(t)] \big[-A + Bp_1(t)\big] p_1(t) \, . 
\label{prison}
\end{equation}
This directly follows from Eq. (\ref{prison0}) of Appendix \ref{AAA}, when the abbreviations   
\begin{equation}
A=P-S  \qquad \mbox{and} \qquad B = P+R-S-T 
\label{pardef}
\end{equation}
are used to pronounce the equation's structure. Setting $dp_1(t)/dt = 0$, one obviously finds three stationary solutions $p_1(t) = p_1^k$, namely
\begin{equation}
p_1^1 = 0\, , \quad p_1^2 = 1 \, \quad \mbox{and} \quad p_1^3 = \frac{A}{B} \, .
\label{statso}
\end{equation}
Not all of these solutions are stable with respect to small deviations. In fact, a linear stability analysis (see Appendix \ref{AAA}) delivers the eigenvalues
\begin{equation}
\lambda_1 = -A, \qquad \lambda_2 = A-B, \qquad \mbox{and} \qquad \lambda_3 = A\left( 1 - \frac{A}{B}\right) . 
\end{equation}
The stationary solution $p_1^k$ is stable with respect to perturbations (i.e. small deviations from them), if $\lambda_k \le 0$, while for $\lambda_k>0$, the deviation will grow in time. 
Due to 
\begin{equation}
A-B=T-R >0 \, , 
\end{equation}
the solution $p_1^2=1$  corresponding to 100\% cooperators is unstable with respect to perturbations, i.e. it will not persist. Moreover, the solution $p_1^1$ will be stable in the prisoner's dilemma because of $A = P-S>0$. This corresponds to 0\% cooperators and 100\% defectors, which agrees with the expected result for the one-shot prisoner's dilemma (if individuals decide according to rational choice theory). In the snowdrift game, however, the stationary solution $p_1^1 = 0$ is unstable due to $A = P-S< 0$, while the additional stationary solution $p_1^3 = A/B< 1$ is stable. Hence, in the snowdrift game with $B< A < 0$, we expect the establishment of a fraction $A/B$ of cooperators. For the prisoner's dilemma, the solution $p_1^3$ does not exist, as it does not fall into the range between 0 and 1 that is required from probabilities: If $B>0$, we have $p_1^3 = A/B > 1$, while $p_1^3 = A/B < 0$ for $B<0$.
\par
In summary, for the prisoner's dilemma, there is no evolutionarily stable solution with a {\it finite} percentage of cooperators, if we do not consider spontaneous strategy mutations (and neglect the effect of spatial correlations through the applied factorization assumption). According to the above, cooperation in the PD is essentially expected to disappear. Strategy mutations, of course, can increase the stable level $p_1^1$ of cooperation from zero to a finite value. Specifically, $p_1^1$ will assume a value close to zero for small values of $r$, while it will converge to $q$ in the limit $r\rightarrow \infty$.
\par
In the next section, we will show that
\begin{enumerate}
\item when the proportions of cooperators and defectors are allowed to vary in space, i.e. if the distribution of cooperators and defectors is {\it inhomogeneous}, the proportion of cooperators may locally grow,
\item we obtain such a variation in space by success-driven motion, as it can {\it de}sta\-bilize a homogeneous distribution of strategies, which gives rise to spatial pattern formation in the population (agglomeration or segregation or both \cite{EPL}).
\end{enumerate}
Together with the well-known fact that a clustering of cooperators can promote cooperation \cite{Pepper,cluster1,cluster2,cluster3}, pattern formation can potentially amplify the level of cooperation, as was demonstrated numerically for a somewhat related model in Ref. \cite{ACS5}.\footnote{In contrast to this EPJB paper, the one published in Advances in Complex Systems (ACS) studies the dynamics in a two-dimensional grid, assuming spatial exclusion (i.e. a cell can only be occupied once), neglecting effects of noise and diffusion, and choosing the payoffs $P=S=0$, which restricts the results to a degenerate case of the prisoner's dilemma and the snowdrift game. Moreover, this EPJB paper focusses on the pattern formation instability rather than the amplification of the level of cooperation,  formalizes social forces resulting from success-driven motion, and discusses pattern formation in the spatial prisoner's dilemma induced by asymmetrical diffusion.}
We will also show that, in contrast to success-driven motion, random motion (``diffusion'' in space) stabilizes the stationary solution $p_1^1$ with 0\% cooperators (or, in the presence of strategy mutations, with a small percentage of cooperators).

\section{Taking into Account Success-Driven Motion and Diffusion in Space}\label{withSpace}

We now assume that individuals are distributed over different locations $x\in [0,L]$ of a one-dimensional space. A generalization to multi-dimensional spaces is easily possible. For the sake of simplicity, we assume that the spatial variable $x$ is scaled by the spatial extension $L$, so that $x$ is dimensionless and varies between 0 and 1. In the following, the proportion of individuals using strategy $i$ at time $t$ and at a location between $x$  and $x+dx$ is represented by $p_i(x,t)dx$ with $p_i(x,t) \ge 0$. Due to the spatial degrees of freedom, the proportion of defectors is not immediately given by the the proportion of cooperators anymore, 
and the previous normalization condition $p_2(t) = 1-p_1(t)$ is replaced by the less restrictive condition 
\begin{equation}
\sum_{i=1}^2 \int\limits_0^L dx \, p_i(x,t) = 1 \, .
\label{normalize2}
\end{equation}
This allows the fractions of cooperators and defectors to uncouple locally, i.e. the proportion of cooperators does not have to decrease anymore by the same amount as the proportion of defectors increases.
\par
Note that, if $\rho_i(x,t) = p_i(x,t)N/L$ represents the \textit{density} of individuals pursuing strategy $i$ at location $x$ and time $t$, Eq. (\ref{normalize2}) can be transferred into the form
\begin{equation}
\sum_{i=1}^2 \int\limits_0^1 dx \, \rho_i(x,t) = \frac{N}{L} = \rho \, ,
\end{equation}
where $N$ is the total number of individuals in the system and $\rho$ their average density.
\par
One may also consider to treat unoccupied space formally like a third strategy $i=0$. In this case, however, the probabilities $p_i(x,t)$ in all locations $x$ add up to the maximum concentration $p_{\rm max}$, see Eq. (\ref{Prefa}). This means
\begin{equation}
p_0(x,t) = p_{\rm max} - p_1(x,t) - p_2(x,t) 
\end{equation}
and
\begin{equation}
\frac{\partial p_0(x,t)}{\partial t} = - \frac{\partial p_1(x,t)}{\partial t} - \frac{\partial p_2(x,t)}{\partial t} \, .	 
\end{equation}
Therefore, $p_0(x,t)$ can be eliminated from the system of equations, because unoccupied space does not {\it interact} with strategies 1 and 2. As a consequence, the dynamics in spatial games with success-driven motion is different from the cyclic dynamics in games considering volunteering \cite{volun}: In order to survive invasion attempts by defectors in the prisoner's dilemma, one could think that cooperators would seek separated locations, where they would be ``loners''. However, cooperators do not tend to maneuver themselves into non-interactive states \cite{ACS5}: On the contrary: The survival of cooperators rather requires to have a larger average number of interaction partners than defectors have. 
\par
After this introductory discussion, let us now extend the game-dynamical equations according to
\begin{eqnarray}
 \frac{\partial p_i(x,t)}{\partial t}  &=&  
 \bigg( \sum_{l=1}^2 p_l \sum_{j=1}^2 P_{ij} p_j
 - \sum_{l=1}^2 \sum_{j=1}^2 p_l P_{lj} p_j \bigg) p_i(x,t) \nonumber \\
 &-& \frac{\partial}{\partial x} \left( p_i(x,t) \frac{\partial E_i(x,t)}{\partial x} \right) 
 + D_i \frac{\partial^2 p_i(x,t)}{\partial x^2} \qquad 
  \label{spacedyn}
\end{eqnarray}
with the local expected success
\begin{equation}
 E_i(x,t) = \sum_{j=1}^2 P_{ij} p_j(x,t) \, ,
\end{equation}
compare Eq. (\ref{compeq}). The additional sum $\sum_l p_l$ had to be introduced for reasons of normalization, as we do not have $p_1+p_2 = 1$ any longer.\footnote{Rather than multiplying the first sum over $j$ in Eq. (\ref{spacedyn}) by $\sum_l p_l$, one could also divide the second sum over $j$  by $\sum_l p_l$, corresponding to a subtraction of  the average expected success $\sum_l p_l E_l/\sum_l p_l$ from the expected success $E_i$. The alternative specification chosen here assumes that the number of strategic game-theoretical interactions of an individual per unit time is proportional to the number of individuals it may interact with, i.e. proportional to $\sum_l p_l(x,t)$. Both specifications are consistent with the game-dynamical equation (\ref{gamedyn}), where $\sum_l p_l(t) = 1$.}
$\partial p_i(x,t)/\partial t$ represents the (partial) time derivative. The first term  in large brackets on the right-hand side of Eq. (\ref{spacedyn}) assumes that locally, an imitation of more successful strategies occurs. An extension of the model to interactions with {\it neighboring} locations would be easily possible. The second term, which depends on $E_i(x,t)$, describes success-driven motion \cite{withTamas,withTadek}. Finally, the last term represents diffusion, and $D_i \ge 0$ are called diffusion coefficients or diffusivities. These terms can be generalized to multi-dimensional spaces by replacing the spatial derivative $\partial/\partial x$ by the nabla operator $\vec{\nabla}$.
\par
The notion of success-driven motion is justified for the following reason: Comparing the term describing success-driven motion with a Fokker-Planck equation \cite{FPG}, one can conclude that it corresponds to a systematic drift with speed
\begin{equation}
V_i(x,t) = \frac{\partial E_i(x,t)}{\partial x} \, . 
\label{expli}
\end{equation}
According to this, individuals move into the direction of the gradient of the expected payoff, i.e. the direction of the (greatest) increase of  $E_i(x,t)$.  In order to take into account capacity constraints (saturation effects), one could introduce a prefactor 
\begin{equation}
 C(x,t) = 1 - \sum_{l=1}^2 \frac{p_l(x,t)N}{\rho_{\rm max}L} \ge 0 \, ,
 \label{Prefa}
\end{equation}
where $\rho_{\rm max} = p_{\rm max} N/L \ge N/L > 0$ represents the maximum density of individuals. This would have to be done in the imitation-based replicator terms in the first line of Eq. (\ref{spacedyn}) as well. In the following, however, we will focus on the case $C=1$, which allows for a local accumulation of individuals. 
\par
The last term in Eq. (\ref{spacedyn}) is a diffusion term which reflects effects of random motion in space \cite{FPG}. It can be easily seen that, for $D_i > 0$, the diffusion term has a smoothing effect: It eventually reduces the proportion $p_i(x,t)$ in places $x$ where the second spatial derivative $\partial^2 p_i/\partial x^2$ is negative, in particular in places  $x$ where the distribution $p_i(x,t)$ has maxima in space. In contrast, the proportion $p_i(x,t)$ increases in time, where $\partial^2 p_i/\partial x^2 > 0$, e.g. where the distribution has its minima. Assuming an additional smoothing term 
\begin{equation}
D_0 \, \frac{\partial^4 p_i(x,t)}{\partial x^4} 
\label{D0}
\end{equation}
with a small constant $D_0 > 0$ on the right-hand side of Eq. (\ref{spacedyn}) makes the numerical solution of this model well-behaved (see Appendix \ref{BBB}).
\par
When Eq. (\ref{spacedyn}) is solved,  one may, for example, assume periodic boundary conditions (i.e. a circular space). In this case, we have $p_i(1,t) = p_i(0,t)$ and $\partial^k p_i(1,t)/\partial x^k = \partial^k p_i(0,t)/\partial x^k$, and by means of partial integration, it can be shown that
\begin{equation}
 \frac{\partial }{\partial t} \sum_{i=1}^2 \int\limits_0^1 dx \, p_i(x,t) = 0 \, . 
\end{equation}
Therefore, Eq. (\ref{spacedyn}) fulfils the normalization condition (\ref{normalize2}) at all times, if it is satisfied at $t=0$. Furthermore, it can be shown that $p_i(x,t) \ge 0$ for all times $t$, if this is true at time $t=0$ for all strategies $i$ and locations $x$. 

\subsection{Comparison with Reaction-Diffusion-Advection Equations}\label{COMPA}

It is noteworthy that the extended game-dynamical model (\ref{spacedyn}) has some similarity with reaction-diffusion-advection (RDA) equations. These equations have been developed to describe the dynamics of chemical reactions with spatial gradients, considering the effects of differential flows and diffusion. Specifically, the kinetics of (binary) chemical reactions is reflected by non-linear terms similar to those in the first line on the right-hand side of Eq. (\ref{spacedyn}). The first term in the second line represents advection terms (differential flows), while the last term of Eq. (\ref{spacedyn}) delineates diffusion effects. 
\par
Considering this apparent similarity, what dynamics do we expect? It is known that reaction-diffusion equations can show a Turing instability \cite{Turing1,Turing3} (also without an advection term). Specifically, for a chemical activator-inhibitor system, one can find a linearly unstable dynamics, if the diffusivities $D_1$ and $D_2$ are different. As a consequence, the concentration of chemicals in space will be non-homogeneous. This effect has been used to explain pattern formation processes in morphogenesis \cite{Meinhardt,Turing2}.
Besides the Turing instability, a second pattern-forming instability can occur when chemical reactions are coupled with differential flows. These so-called ``differential flow-induced chemical instabilities'' (DIFICI) can occur even for equal or vanishing diffusivities $D_i = D$, but they may also interact with the Turing instability \cite{DIF1,DIF2,DIF3,DIF4,DIF5}.
\par
The difference of the extended game-dynamical equation (\ref{spacedyn}) as compared to the RDA equations lies in the specification of the velocity $V_i(x,t)$, which is determined by the gradient of the expected success $E_i(x,t)$. Therefore,  the advective term is self-generated by the success of the players. We may rewrite the corresponding term in Eq. (\ref{spacedyn}) as follows:
\begin{equation}
- \frac{\partial}{\partial x} \left( p_i \frac{\partial E_i}{\partial x} \right) 
= - \sum_{j=1}^2 P_{ij}  \frac{\partial p_i}{\partial x} \frac{\partial p_j}{\partial x} 
-  \sum_{j=1}^2 p_i P_{ij} \frac{\partial^2 p_j}{\partial x^2} \, . \label{CONSI}
\end{equation}
According to the last term of this equation, the advection term related to success-driven motion implies effects similar to a diffusion term with negative diffusion coefficients $p_i P_{ij}$ (if $P_{ij} > 0$). Additionally, there is a non-linear dependence on the gradients $\partial p_i/\partial x$ and $\partial p_j/\partial x$ of the strategy distributions in space. Both terms couple the dynamics of different strategies $j$. 
\par
As a consequence of this, the resulting instability conditions and dynamics are different from the RDA equations. We will show that, without success-driven motion, diffusion cannot trigger pattern formation. Diffusion rather counteracts spatial inhomogeneities. Success-driven motion, in contrast, causes pattern formation in a large area of the parameter space of payoffs, partly because of a negative diffusion effect. \par
Besides this, we will show that different diffusivities $D_i$ may trigger a pattern formation instability in case of payoff parameters, for which success-driven motion does not destabilize homogeneous strategy distributions in the absence of diffusion. This counter-intuitive effect reminds of the Turing instability, although the underlying mathematical model is different, as pointed out before. In particular, the largest growth rate does not occur for finite wave numbers, as Appendix \ref{BBB} shows. The instability for asymmetric diffusion is rather related to the problem of  ``noise-induced transitions'' in systems with multiplicative noise, which are characterized by space-dependent diffusion coefficients  \cite{Horsthemke}. A further interesting aspect of success-driven motion is the circumstance that the space-dependent diffusion effects go back to binary interactions, as the multiplicative dependence on $p_i$ and $p_j$ shows.

\section{Linear Instability of the Prisoner's Dilemma and the Snowdrift Game} \label{results}

After inserting the payoffs (\ref{PAYOFF}) of the prisoner's dilemma and the snowdrift game into Eq. (\ref{spacedyn}), one can see the favorable effect on the spreading of cooperation that the spatial dependence, in particular the  relaxed normalization condition (\ref{normalize2}) can have: For $i=1$, Eq. (\ref{spacedyn}) becomes
\begin{eqnarray}
 \frac{\partial p_1(x,t)}{\partial t}  &=&  
 p_2 \Big[ (R - T) p_1 + (S - P) p_2 \Big] p_1 \nonumber \\
 &+& \frac{\partial}{\partial x} \left[ (D_1 - p_1 R) \frac{\partial p_1}{\partial x} - p_1 S \frac{\partial p_2}{\partial x}\right]  \, . \qquad 
  \label{spacedyn4}
\end{eqnarray}
It is now an interesting question, whether the agglomeration of cooperators can be supported by success-driven motion. In fact, the second line of Eq. (\ref{spacedyn4}) can be rewritten as
\begin{equation}
(D_1 - p_1R)  \frac{\partial^2 p_1}{\partial x^2} - R \left(\frac{\partial p_1}{\partial x}\right)^2 \! 
- p_1S  \frac{\partial^2 p_2}{\partial x^2} - S \frac{\partial p_1}{\partial x} \, \frac{\partial p_2}{\partial x}  \, . 
  \label{spacedyn5}
\end{equation}
This shows that a curvature $\partial^2 p_i/\partial x^2 < 0$ can support the increase of the proportion of strategy $i$ as compared to the game-dynamical equation (\ref{gamedyn}) without spatial dependence. 
The situation becomes even clearer, if a linear stability analysis of Eqs. (\ref{spacedyn4}) is performed (see Appendix \ref{BBB}). The result is as follows: If the square of the wave number $\kappa$, which relates to the curvature of the strategy distribution, is large enough (i.e. if the related cluster size is sufficiently small), the replicator terms in the first line on the right-hand side of equation (\ref{spacedyn}) 
become negligible. Therefore, the conditions, 
under which homogeneous initial strategy distributions $p_i^0$ are linearly unstable, simplify to 
\begin{eqnarray}
p_1^0R + p_2^0P > D_1 + D_2 
\label{so1}
\end{eqnarray}
and
\begin{equation}
(p_1^0R-D_1) (p_2^0P-D_2) < p_1^0 p_2^0 ST  \, ,
\label{so}
\end{equation}
see Eqs. (\ref{directed1}) and (\ref{directed}).\footnote{This instability condition has been studied in the context of success-driven motion {\it without} imitation and for games with symmetrical payoff matrices (i.e. $P_{ij} = P_{ji}$), which show a particular behavior \cite{EPL}. Over here, in contrast, we investigate continuous spatial games involving imitation (selection of more successful strategies), and focus on {\it asymmetrical} games such as the prisoner's dilemma and the snowdrift game (see Sec. \ref{results}), which behave very differently.} 
\par 
If condition (\ref{so1}) or condition (\ref{so}) is fulfilled, we expect emergent spatio-temporal pattern formation, basically agglomeration or segregation or both \cite{EPL}. As has been pointed out before, the agglomeration of cooperators can increase the level of cooperation. This effect is not possible, when spatial interactions are neglected. In that case, we stay with Eq. (\ref{prison}), and the favorable pattern-formation effect cannot occur. 
\par
\begin{figure}[htbp]
\begin{center}
\includegraphics[width=9cm]{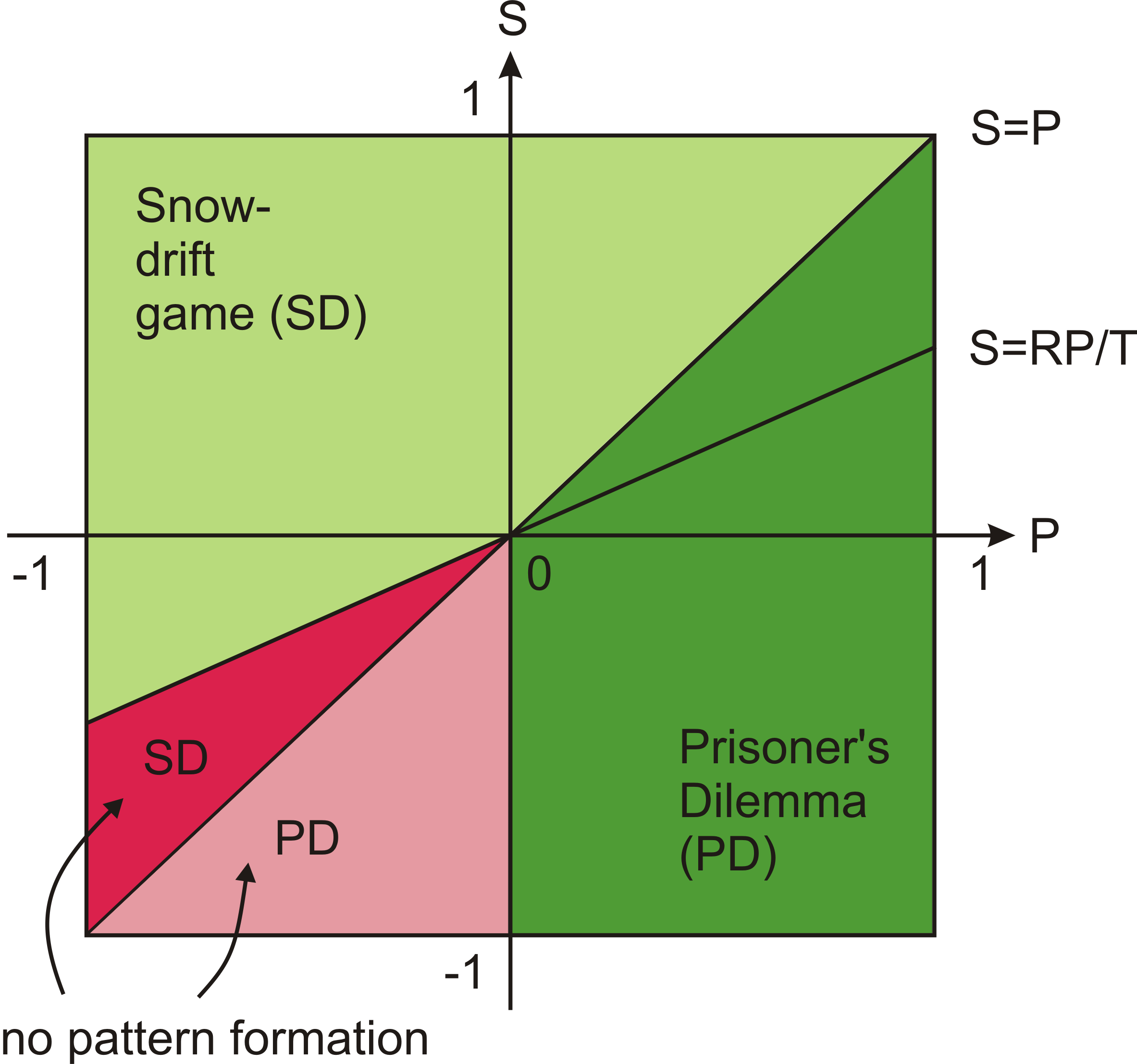}
\end{center}
\caption[]{Payoff-dependence of pattern-formation in the prisoner's dilemma with $S<P$ and the snowdrift game with $S>P$ according to a linear stability analysis for spatial games with success-driven motion, but no diffusion, strategy mutations, or noise. One can clearly see that spontaneous pattern-formation prevails (green area), and that there is only a small area for $P<0$ (marked red), where a homogeneous initial condition is stable with respect to small perturbations.}
\label{FIG1}
\end{figure}

Let us now discuss a variety of different cases:
\begin{enumerate}
\item In case of diffusive motion only (i.e. no success-driven motion), Eq. (\ref{so1}) must be replaced by
\begin{equation}
0 > D_1 + D_2 
\end{equation}
and Eq. (\ref{so}) by
\begin{equation}
D_1D_2 < 0 \, , 
\end{equation}
which cannot be fulfilled. Therefore, diffusion without success-driven motion does not support pattern formation or a related increase in the level of cooperation.
\item In the case of success-driven motion with finite diffusion,
Eqs. (\ref{so1}) and (\ref{so}) imply 
\begin{eqnarray}
p_1^0 R + (1-p_1^0) P  > D_1 + D_2 
\label{so3}
\end{eqnarray}
and
\begin{equation}
 (p_1^0R-D_1) \big[ (1-p_1^0)P-D_2\big]
< p_1^0 (1-p_1^0) ST  \, ,
\label{so4}
\end{equation}
if the normalization condition $p_2^0 = 1 - p_1^0$ for a homogeneous initial condition is taken into account. These instability conditions hold for both, the prisoner's dilemma and the snowdrift game.  
Inequality (\ref{so3}) basically says that, in order to find spontaneous pattern formation, the agglomerative tendency $R$ of cooperators or the agglomerative tendency $P$ of defectors (or both) must be larger than the diffusive tendency. This agglomeration, of course, requires the reward $R$ of cooperation to be positive, otherwise cooperators would not like to stay in the same location. The alternative instability condition (\ref{so4}) requires that the product $RP$ of the payoffs resulting when individuals of the same strategy meet each other is smaller than the product $ST$ of payoffs resulting when individuals with different strategies meet each other. This basically excludes a coexistence of the two strategies in the same location and is expected to cause segregation. It is noteworthy that condition (\ref{so4}) is not invariant with respect to shifts of all payoffs $P_{ij}$ by  a constant value $c$, in contrast to the replicator equations (\ref{prison0}).
\item In case of the prisoner's dilemma without strategy mutations, the stable stationary solution for the case without spontaneous strategy changes is $p_1^0 = p_1^1 = 0$. This simplifies the instability conditions further, yielding
\begin{eqnarray}
P   > D_1 + D_2 
\label{so5}
\end{eqnarray}
and
\begin{equation}
-D_1 (P-D_2) < 0  \, .
\label{so6}
\end{equation}
In order to fulfil one of these conditions, the punishment $P$ must be positive and larger than $D_2$ to support pattern formation (here: an aggregation of defectors). Naturally, the survival or spreading of cooperators requires the initial existence of a finite proportion $p_1^0 > 0$ of them, see the previous case. 
\item In the special case $P=S=0$ \cite{ACS5}, 
Eqs. (\ref{so3}) and (\ref{so4}) become
\begin{equation}
p_1^0R  > D_1 + D_2 
\end{equation}
and
\begin{equation}
- (p_1^0R-D_1) D_2 < 0 \, ,  
\end{equation}
which requires $p_1^0R > D_1$. Therefore, a finite initial proportion $p_1^0$ of cooperators is needed again for pattern formation (an agglomeration of cooperators). This can be easily reached by spontaneous strategy changes.
\item Neglecting diffusion for a moment (i.e. setting $D_1=D_2 =0$), 
{\it no} pattern formation should occur, if the condition
\begin{equation}
RP > ST 
 \label{ineq1}
\end{equation}
and, at the same time,
\begin{equation}
p_1^0R + (1-p_1^0)P < 0  
\label{ineq2}
\end{equation}
is fulfilled. Equation (\ref{ineq1}) implies the stability condition 
\begin{equation}
S < \frac{RP}{T}  \, . 
\label{SRT}
\end{equation}
Besides $T>R$, we have to consider here that $S<P$ in the prisoner's dilemma and $S>P$ in the snowdrift game (see Fig. \ref{FIG1}).\footnote{Strictly speaking, we also need to take into account Eq. (\ref{ineq2}), which implies $P<0$ for the stationary solution $p_1^0 = p_1^1=0$ of the prisoner's dilemma and generally $P < - p_1^0 R/(1-p_1^0)$. 
In case of the snowdrift game \cite{Space3},  it is adequate to insert the stationary solution $p_1^0 = p_1^3 = A/B$, which is stable in case of {\it no} spatial interactions. This leads to the condition 
$P < - p_1^3R/(1-p_1^3) = AR/(A-B) = (P-S)R/(T-R)$, i.e. 
$SR < (2R- T) P$. The question is, whether this condition will reduce the previously determined area of stability given by $S<RP/T$ with $P<0$, see Eq. (\ref{SRT}) and Fig. \ref{FIG1}. This would be the case, if $(2-T/R)P>RP/T$, which by multiplication with $RT/P$ becomes
$(2RT-T^2) > R^2$ or $(T-R)^2 < 0$. Since this condition cannot be fulfilled, it does not impose any further restrictions on the stability area in the snowdrift game.}
\item Finally, let us  assume that the stability conditions
$ST-RP < 0$ and $p_1^0 R+ (1-p_1^0)P<0$ for the previous case without diffusion ($D_1=D_2=0$) are {\it fulfilled}, so that no patterns will emerge. Then, depending on the parameter values, the instability condition following from Eq. (\ref{so4}),
\begin{eqnarray}
& &  -(D_1-D_2) \underbrace{(1-p_1^0)}_{\ge 0}P \nonumber \\ 
&<& \underbrace{D_2}_{\ge 0} \underbrace{[p_1^0R+ (1-p_1^0)P]}_{<0} \nonumber \\
 & & \underbrace{-D_1D_2}_{\le 0} + \underbrace{p_1^0(1-p_1^0)}_{\ge 0} \underbrace{(ST-RP)}_{<0} \, , \qquad
\end{eqnarray}
may still be matched, if  $(D_1-D_2)$ is sufficiently large. Therefore, aymmetrical diffusion ($D_1\ne D_2$) can trigger a pattern formation instability, where the spatio-temporal strategy distribution without diffusion would be stable. The situation is clearly different for symmetrical diffusion with $D_1=D_2$, which cannot support pattern formation. 
\par
Although the instability due to asymmetrical diffusion reminds of the Turing instability, it must be distinguished from it (see Sec. \ref{COMPA}). So, how can the instability then be explained? The reason for it may be imagined as follows: In order to survive and spread, cooperators need to be able to agglomerate locally and to invade new locations. While the first requirement is supported by success-driven motion, the last one is promoted by a larger diffusivity $D_1 > D_2$ of cooperators. 
\end{enumerate}

\section{Social Forces in Spatial Games with Success-Driven Motion}\label{forces}

In the previous section, we have shown how success-driven motion destabilizes homogeneous strategy distributions in space. This analysis was based on the study of linear (in)stability (see Appendix \ref{BBB}). But what happens, when the deviation from the homogeneous strategy distribution is large, i.e. the gradients $\partial p_i/\partial x$ are not negligible any longer? This can be answered by writing Eq. (\ref{expli}) explicitly, which becomes
\begin{equation}
V_i(x,t) = \sum_{j=1}^2 P_{ij} \frac{\partial p_j(x,t)}{\partial x} = \sum_{j=1}^2 f_{ij}(x,t) \, .
\end{equation}
Here, the expression
\begin{equation}
f_{ij}(x,t) =  P_{ij} \frac{\partial p_j(x,t)}{\partial x} 
\end{equation}
(which can be extended by saturation effects),
may be interpreted as interaction force (``social force'') excerted by individuals using strategy $j$ on an individual using strategy $i$.\footnote{Note that this identification of a speed with a force is sometimes used for dissipative motion of the kind $m_\alpha d^2x_\alpha/dt^2 = - \gamma_\alpha dx_\alpha/dt + \sum_\beta F_{\alpha\beta}(t)$, where $x_\alpha(t)$ is the location of an individual $\alpha$, the ``mass'' $m_\alpha$ reflects inertia, $\gamma_\alpha$ is a friction coefficient, and $F_{\alpha\beta}(t)$ are interaction forces. In the limiting case $m_\alpha \rightarrow 0$, we can make the adiabatic approximation $dx_\alpha/dt = \sum_\beta F_{\alpha\beta}(t)/\gamma_\alpha = \sum_\beta f_{\alpha\beta}(t)$, where
$dx_\alpha/dt$ is a speed and $f_{\alpha\beta}(t)$ are proportional to the interaction forces $F_{\alpha\beta}(t)$. Hence, the quantities $f_{\alpha\beta}(t)$ are sometimes called ``forces'' themselves.}
It is visible that the {\it sign} of $P_{ij}$ determines the {\it character} of the force. The force is attractive for positive payoffs $P_{ij}>0$ and repulsive for negative payoffs $P_{ij}<0$. The
{\it direction} of the force, however, is determined by spatial changes $\partial p_j(x,t)/\partial x$ in the strategy distribution $p_j(x,t)$ (i.e. not by the strategy distribution itself).  
\par
It is not the {\it size} of the payoffs $P_{ij}$ which determines the strength of the interaction force, but the payoff {\it times} the gradient of the distribution of the strategy $j$ one interacts with (and the availability and reachability of more favorable neighboring locations, if the saturation prefactor $C$ is taken into account).  Due to the dependence on the gradient $\partial p_j(x,t)/\partial x$, the impact of a dispersed strategy $j$ on individuals using strategy $i$ is negligible. This particularly applies to scenarios with negative self-interactions ($P_{jj} < 0$). 
\par
Note that success-driven motion may be caused by repulsion away from the current location or by attraction towards more favorable neighborhoods. In the prisoner's dilemma, for example, cooperators and defectors feel a strong attraction towards areas with a higher proportion of cooperators. However, cooperators seek each other mutually, while the attraction between defectors and cooperators is weaker. This is due to $T+S < 2R$, see inequality (\ref{xcv}). As a result, even if $P>0$, cooperators are moving away from defectors due to $R>P$ in order to find more cooperative locations, while defectors are following them.  
\par
Another interesting case is the game with the payoffs $P_{11}=P_{22}=-P$ and $P_{12}=P_{21}=Q>P$, where we have negative self-interactions among identical strategies and positive interactions between different strategies. Simulations for the no-imitation case show that, despite of the dispersive tendency of each strategy, strategies tend to agglomerate in certain locations thanks to the stronger attractive interactions between different strategies (see Fig. 3 in Ref. \cite{withTadek}).
\par
The idea of social forces is long-standing. Montroll used the term to explain logistic growth laws \cite{Montroll}, and Lewin introduced the concept of social fields to the social sciences in analogy to electrical fields in physics \cite{Lewin}. However, a formalization of a widely applicable social force concept was missing for a long time. In the meantime, social forces were successfully used to describe the dynamics of interacting vehicles \cite{tilch} or pedestrians \cite{molnar}, but there, the attractive or repulsive nature was just assumed. Attempts to systematically derive social forces  from an underlying decision mechanism were based on direct pair interactions in behavioral spaces (e.g. opinion spaces), with the observation that imitative interactions or the readiness for compromises had attrative effects \cite{Kluwer,MatSoc}. Here, for the first time, we present a formulation of social forces in game-theoretical terms. Considering the great variety of different games, depending on the respective specification of the payoffs $P_{ij}$, this is expected to find a wide range of applications,
in particular as success-driven motion has been found to produce interesting and relevant
pattern formation phenomena \cite{withTadek,ACS5}.

\section{Summary, Discussion, and Outlook}\label{summa}

In this paper, we have started from the game-dynamical equations (replicator equation), which can be derived from imitative pair interactions between individuals \cite{imit1,imit2}. It has been shown that no cooperation is expected in the prisoner's dilemma, if no spontaneous strategy mutations are taken into account, otherwise there will be a significant, but usually low level of cooperation. In the snowdrift game, in contrast, the stationary solution corresponding to no cooperation is unstable, and there is a stable solution with a finite level of cooperation.
\par
These considerations have been carried out to illustrate the major difference that the introduction of spatial interactions based on success-driven motion and diffusion makes. While diffusion itself tends to support homogeneous strategy distributions rather than pattern formation, success-driven motion implies an unstable spatio-temporal dynamics under a wide range of conditions.
As a consequence, small fluctations can destabilize a homogeneous distribution of strategies. Under such conditions, the formation of emergent patterns (agglomeration, segregation, or both) is expected. 
The resulting dynamics may be understood in terms of social forces, which have been formulated here in game-theoretical terms. 
\par
The destabilization of homogeneous strategy distributions and the related occurence of spontaneous pattern formation has, for example, a great importance for the survival and spreading of cooperators in the prisoner's dilemma. While this has been studied numerically in the past \cite{ACS5}, future work based on the model of this paper and extensions of it shall analytically study conditions for the promotion of cooperation. For example, it will be interesting to investigate, how relevant the imitation of strategies in {\it neighboring} locations is, how important is the {\it rate} of strategy changes as compared to location changes, and how crucial is a territorial effect (i.e. a limitation $p_{\rm max}$ of the local concentration of individuals, which may protect cooperators from invasion by defectors).
\par
Of course, instead of studying the continuous game-dynamical model with success-driven motion and calculating its instability conditions, one can also perform agent-based simulations for a discretized version of the model. For the case without imitation of superior strategies and symmetrical payoffs ($P_{ij}=P_{ji}$), it has been shown that the analytical instability conditions surprisingly well predict the parameter areas of the agent-based model, where pattern-formation takes place \cite{EPL}. Despite the difference in the previously studied model (see footnote 2), this is also expected to be true for the non-symmetrical games studied here, in particular as we found that the influence of imitation on the instability condition is negligible, if the wave number $\kappa$ characterizing inhomogeneities in the initial distribution is large. This simplified the stability analysis a lot. Moreover, it was shown that {\it asymmetrical} diffusion can drive our game-theoretical model with success-driven motion from the stable regime into the unstable regime. While this reminds of the Turing instability \cite{Turing1}, it is actually different from it: Compared to reaction-diffusion-advection equations, the equations underlying the game-dynamical model with success-driven motion belong to another mathematical class, as is elaborated in Sec. \ref{COMPA}.
\par
In Sec. \ref{forces}, it was pointed out that, in the prisoner's dilemma, cooperators evade defectors, who seek cooperators. Therefore, some effects of success-driven motion (leaving unfavorable neighborhoods) may be interpreted as punishment of the previous interaction partners, who are left behind with a lower overall payoff. However, ``movement punishment'' of defectors by leaving unfavorable environments is different from the ``costly'' or ``altruistic punishment'' discussed in the literature \cite{Punish}: In the strict sense, success-driven motion neither imposes costs on a moving individual nor on the previous interaction partners. If we would introduce a cost of movement, it would have to be paid by both,  cooperators who evade defectors, and defectors who follow them. Therefore, costly motion would be expected to yield similar results as before, but it would still be different from altruistic punishment. It should also be pronounced that, besides avoiding unfavorable locations, success-driven motion implies the seeking of favorable environments, which has nothing to do with punishment. Without this element, e.g. when individuals leave unfavorable locations based on a random, diffusive motion, success-driven motion is not effective in promoting cooperation. Therefore, the mechanism of success-driven motion, despite some similar features, is clearly to be distinguished from the mechanism of punishment. 
\par
Finally, note that migration may be considered as one realization of success-driven motion. Before, the statistics of migration behavior was modeled by the gravity law \cite{gravity1,gravity2} or entropy approaches \cite{entropy1,entropy2}, while its dynamics was described by partial differential equations \cite{part1,part2} and models from statistical physics \cite{Weidlich}.  The particular potential of the approach proposed in this paper lies in the integration of migration into a game-theoretical framework, as we formalize success-driven motion in terms of payoffs and strategy distributions in space and time. Such integrated approaches are needed in the social sciences to allow for consistent interpretations of empirical findings within a unified framework.

\subsection*{Acknowledgments} 
The author would like to thank Peter Felten for preparing Fig. 1 and Christoph Hauert for comments on manuscript \cite{ACS5}.

\appendix

\section{Linear Stability Analysis of the Game-Dynamical Equation Without Spatial Interactions}\label{AAA}

Inserting the payoffs (\ref{PAYOFF}) of the prisoner's dilemma or the snowdrift game into Eq. (\ref{gamedyn}), we get the game-dynamical equation
\begin{eqnarray}
\frac{dp_1(t)}{dt} &=& \big[ Rp_1 + Sp_2 - R(p_1)^2 \nonumber \\
&-& (S+T)p_1p_2 - P(p_2)^2 \big] p_1(t) \, . 
\end{eqnarray}
Considering Eq. (\ref{normalize}), i.e. $p_2(t) = 1 - p_1(t)$, we find
\begin{eqnarray}
\frac{dp_1}{dt} &=& \big[ Rp_1 + S(1-p_1) - R(p_1)^2 \nonumber \\
&& -(S+T)p_1(1-p_1) - P(1-p_1)^2 \big] p_1(t)\nonumber \\[1mm]
&=& (1-p_1) \big[(S-P) + (P+R-S-T)p_1\big] p_1(t) \, .\qquad
\label{prison0}
\end{eqnarray} 
This is a mean-value equation, which assumes a factorization of joint probabilities, i.e. it neglects correlations \cite{Kluwer}. Nevertheless, the following analysis is suited to provide insights into the dynamics of the prisoner's dilemma and the snowdrift game. 
Introducing the useful abbreviations
\begin{equation}
A=P-S  \qquad \mbox{and} \qquad B = P+R-S-T ,
\end{equation}
Eq. (\ref{prison0}) can be further simplified, and we get
\begin{equation}
\frac{dp_1(t)}{dt} = [1-p_1(t)] \big[-A + Bp_1(t)\big] p_1(t) \, .
\end{equation}   
Obviously, shifting all payoffs $P_{ij}$ by a constant value $c$ does not change Eq. (\ref{prison}),
in contrast to the case involving spatial interactions discussed later.
\par
Let $p_1^k$ with $k\in \{1,2,3\}$ denote the stationary solutions (\ref{statso}) of Eq. (\ref{prison}), defined by the requirement $dp_1/dt = 0$. In order to analyze the stability of these solutions with respect to small deviations 
\begin{equation}
\delta p_1(t) = p_1(t) - p_1^k \, , \label{devi}
\end{equation}
we perform a linear stability analysis in the following. For this, we insert Eq. (\ref{devi}) into Eq. \ref{prison}), which yields
\begin{eqnarray}
\frac{d\delta p_1(t)}{dt} &=& (1-p_1^k - \delta p_1) 
(-A + Bp_1^k + B\, \delta p_1)(p_1^k + \delta p_1) \nonumber \\
&=& \big[ (1-p_1^k)p_1^k  + (1-2p_1^k)\delta p_1  - (\delta p_1)^2 \big] \nonumber \\
&\times& (-A + Bp_1^k + B\, \delta p_1) \, .
\end{eqnarray}
If we concentrate on sufficiently small deviations $\delta p_1(t)$, terms containing factors $[\delta p_1(t)]^m$ with an integer exponent $m> 1$ can be considered much smaller than terms containing a factor $\delta p_1(t)$. Therefore, we may linearize the above equations by dropping higher-order terms proportionally to  $[\delta p_1(t)]^m$ with $m>1$. This gives
\begin{eqnarray}
\frac{d\delta p_1}{dt} 
&=&  \big[ (1-p_1^k)p_1^k + (1-2p_1^k)\delta p_1(t) \big](-A+Bp_1^k) \nonumber \\
& & + (1-p_1^k)p_1^k B\, \delta p_1(t ) \nonumber \\[1mm]
&=& \big[ (1-2p_1^k) (-A+Bp_1^k) +  (1-p_1^k)p_1^k B\big] \delta p_1(t) ,\qquad  
\label{linearized}
\end{eqnarray}
as $(1-p_1^k)p_1^k(-A+Bp_1^k) = 0$ for all stationary solutions $p_1^k$. With 
the abbreviation 
\begin{equation}
\lambda_k = (1-2p_1^k) (-A+Bp_1^k) +  (1-p_1^k)p_1^k B \, , 
\end{equation}
we can write
\begin{equation}
\frac{d\delta p_1(t)}{dt} = \lambda_k \, \delta p_1(t) \, . 
\end{equation}
If $\lambda_k < 0$, the deviation $\delta p_1(t)$ will exponentially decay with time, i.e. the solution will converge to the stationary solution $p_1^k$, which implies its stability. If $\lambda_k > 0$, however, the deviation will grow in time, and the stationary solution $p_1^k$ is unstable. For the stationary solutions $p_1^1 = 0$, $p_1^2 = 1$, and $p_1^3 = A/B$ given in Eq. (\ref{statso}), 
we can easily find
\begin{equation}
 \lambda_1 = -A,  \quad \lambda_2 = A-B \, , 
 \quad \mbox{and} \quad \lambda_3 = A \left( 1 - \frac{A}{B}\right) \, , 
\end{equation}
respectively.

\section{Linear Stability Analysis of the Model with Success-Driven Motion and Diffusion}\label{BBB}

In order to understand spatio-temporal pattern formation, it is not enough to formulate the (social) interaction forces determining the {\it motion} of individuals. We also need to grasp, why spatial patterns can {\it emerge} from small perturbations, even if the initial distribution of strategies is uniform (homogeneous) in space.
\par
If Eq. (\ref{spacedyn}) is written explicitly for $i=1$, we get
\begin{eqnarray}
 \frac{\partial p_1}{\partial t}  &=&  \Big[ (p_1+p_2)(P_{11}p_1+P_{12}p_2)
 - P_{11}(p_1)^2  \nonumber \\
 & & - (P_{12}+P_{21})p_1p_2 
 - P_{22} (p_2)^2 \Big] p_1(x,t) \nonumber \\
 &-& \frac{\partial}{\partial x} \bigg[ p_1 \bigg( P_{11} \frac{\partial p_1}{\partial x}
   + P_{12} \frac{\partial p_2}{\partial x} \bigg) \bigg] 
 + D_1\frac{\partial^2 p_1}{\partial x^2} \, .\quad  
   \label{spacedyn2}
\end{eqnarray}
The equation for $i=2$ looks identical, if only $p_1(x,t)$ and $p_2(x,t)$ are exchanged in all places, and the same is done with the indices 1 and 2. We will now assume a homogeneous initial condition $p_i(x,0) = p_i^0$ 
(i.e. a uniform distribution of strategies $i$ in space) and study the spatio-temporal evolution of the deviations $\delta p_i(x,t) = p_i(x,t) - p_i^0$. Let us insert for $p_1^0$ one of the values $p_1^k$, which are stationary solutions of the partial differential equation (\ref{spacedyn}), as $p_2^0 = (1-p_1^0)$ holds for homogeneous strategy distributions due to the normalization condition (\ref{normalize2}). Assuming small deviations $\delta p_i(x,t)$ and linearizing Eq. (\ref{spacedyn2}) by neglecting non-linear terms, we obtain
\begin{eqnarray}
 \frac{\partial \delta p_1}{\partial t}  &=&  \Big[ (P_{11}-P_{21}) p_1^0p_2^0 + (P_{12}-P_{22})(p_2^0)^2 \Big] \delta p_1 \nonumber \\
&+& \Big[ (P_{11}-P_{21})(p_1^0\,\delta p_2+p_2^0\,\delta p_1) \nonumber \\
& & + 2(P_{12}-P_{22})p_2^0\,\delta p_2\Big] p_1^0 \nonumber \\
 &-& p_1^0  \bigg( P_{11} \frac{\partial^2 \delta p_1}{\partial x^2} 
 + P_{12} \frac{\partial^2 \delta p_2}{\partial x^2} \bigg) 
 + D_1\frac{\partial^2 \delta p_1}{\partial x^2} \, . \qquad
   \label{linea}
\end{eqnarray}
Again, a mutation term $W_2\delta p_2(x,t) - W_1\delta p_1(x,t)$ reflecting spontaneous strategy changes may be added, see Eq. (\ref{mutant}). The analogous equation for $\delta p_2(x,t)$ is obtained by exchanging strategies $1$ and $2$. 
\par
In Eq. (\ref{linea}), it can be easily seen that success-driven motion with $P_{ij}>0$ has a similar functional form, but the opposite sign as the diffusion term. While the latter causes a homogenization in space, success-driven motion can cause local agglomeration \cite{withTadek}, first of all for $P_{ij} > 0$. 
\par
It is known that linear partial differential equations like Eq. (\ref{linea}) are solved by (a superposition of) functions of the kind 
\begin{equation}
\delta p_i(x,t) = \mbox{e}^{\tilde{\lambda}t}  \Big[a_{i} \cos(\kappa x) + b_{i} \sin(\kappa x )\Big] 
\, ,
\label{ansatz}
\end{equation}
where $a_{i}$ and $b_{i}$ are initial amplitudes,  
$\tilde{\lambda}=\tilde{\lambda}(\kappa)$ is their growth rate (if positive, or a decay rate, if negative),  
and $\kappa = \kappa_n = 2\pi n/L$ with $n \in \{1,2,\dots\}$ are possible ``wave numbers''. The ``wave length'' $2\pi/\kappa = L/n$ may be imagined as the extension of a cluster of strategy $i$ in space. Obviously, possible wave lengths in case of a circular space of diameter $L=1$ are fractions $L/n$. 
The general solution of Eq. (\ref{linea}) is
\begin{equation}
\delta p_i(x,t) = \sum_{n=1}^\infty \mbox{e}^{\tilde{\lambda}(\kappa_n) t}  \Big[a_{i,n} \cos(\kappa_n x) + b_{i,n} \sin(\kappa_n x )\Big] 
\, , \label{generalsol}
\end{equation}
i.e. a linear superposition of solutions of the form (\ref{ansatz}) with all possible wave numbers $\kappa_n$. For $t=0$, the exponential prefactor $\mbox{e}^{\tilde{\lambda}(\kappa_n)t}$ becomes 1,  and Eq. (\ref{generalsol}) may then be viewed as the Fourier series of the spatial dependence of the initial condition $\delta p(i,x,0)$. Hence, the amplitudes $a_{i,n}$ and $b_{i,n}$ correspond to the Euler-Fourier coefficients \cite{Maths}.
\par
Let us now determine the possible eigenvalues $\tilde{\lambda}(\kappa)$.
For the ansatz (\ref{ansatz}), we have $\partial \delta p_i(x,t)/\partial t = \tilde{\lambda} \delta p_i(x,t)$ and $\partial^2 \delta p_i(x,t)/\partial x^2 
= -\kappa^2 \delta p_i(x,t)$. Therefore, the linearized equations can be cast into the following
form of an eigenvalue problem with eigenvalues $\tilde{\lambda}$:
\begin{equation}
\tilde{\lambda} \left(
\begin{array}{c}
\delta p_1(x,t) \\
 \\
\delta p_2(x,t)
\end{array} \right)
= \underbrace{\left(\begin{array}{ccc}
M_{11} & & M_{12} \\  
 & & \\
M_{21} & & M_{22} 
\end{array}\right)}_{=\underline{M}}
\left(
\begin{array}{c}
\delta p_1(x,t) \\
\\
\delta p_2(x,t)
\end{array} \right) \, . 
\label{eigenvalue}
\end{equation}
Here, we have introduced the abbreviations
\begin{eqnarray}
M_{11} &=& A_{11} + (p_1^0 P_{11}-D_1)\kappa^2 \, , \label{abb1} \\
M_{12} &=& A_{12} + p_1^0 P_{12}\kappa^2 \, , \\
M_{21} &=& A_{21} + p_2^0 P_{21}\kappa^2 \, , \\
M_{22} &=& A_{22} + (p_2^0 P_{22}-D_2)\kappa^2 \label{abb4}
\end{eqnarray}
with
\begin{eqnarray}
A_{11} &=&  \Big[ (P_{12}-P_{22})p_2^0 + 2(P_{11}-P_{21})p_1^0\Big] p_2^0  \\
A_{12} &=& \Big[(P_{11}-P_{21})p_1^0 +  2(P_{12}-P_{22})p_2^0\Big] p_1^0 \, , \\
A_{21} &=& \Big[ (P_{22}-P_{12})p_2^0 + 2(P_{21}-P_{11})p_1^0\Big] p_2^0 \, ,  \\
A_{22} &=& \Big[ (P_{21}-P_{11})p_1^0 + 2(P_{22}-P_{12})p_2^0\Big] p_1^0  
\end{eqnarray}
The eigenvalue problem (\ref{eigenvalue}) can only be solved, if the determinant of the matrix $(\underline{M} - \tilde{\lambda}\underline{1})$ vanishes, where $\underline{1}$ denotes the unit matrix
\cite{Maths}.
In other words, $\tilde{\lambda}$ are solutions of the so-called ``characteristic polynomial''
\begin{eqnarray}
& &  (M_{11} - \tilde{\lambda})(M_{22} - \tilde{\lambda}) - M_{12}M_{21}  \nonumber \\
&=& \tilde{\lambda}^2 - (M_{11}+M_{22})\tilde{\lambda} + M_{11}M_{22} - M_{12}M_{21} = 0 \, . \qquad 
\end{eqnarray}
This polynomial is of degree 2 in $\tilde{\lambda}$ and has the following two solutions:
\begin{eqnarray}
\tilde{\lambda}(\kappa) &=& \frac{M_{11}+M_{22}}{2} \nonumber \\
&\pm& \frac{1}{2} \sqrt{ (M_{11}+M_{22})^2 - 4(M_{11}M_{22} - M_{12}M_{21} )}
\, . \qquad \label{EM}
\end{eqnarray}
The fastest growing mode (i.e. the value of the wave number $\kappa$ with the largest real value $\mbox{Re}(\tilde{\lambda})$ of $\tilde{\lambda}$) usually determines the length scale of the emerging patterns. Considering (\ref{abb1}) to (\ref{abb4}), we can easily see that the largest value of $\mbox{Re}(\tilde{\lambda})$ is reached in the limit $\kappa \rightarrow \infty$. This is due to the relationship of success-driven motion with negative diffusion. Hence, the finally resulting distribution would be a superposition of delta peaks. As this is not favorable from a numerical perspective, the smoothing term (\ref{D0}) may be added, which implies the additional terms $D_0\kappa^4$ in Eqs. (\ref{abb1}) and (\ref{abb4}). These terms imply, in fact, that $\mbox{Re}(\tilde{\lambda})$ reaches its maximum value for a finite value of $\kappa$. Note, however, that discrete models involving success-driven motion also tend to end up with distributions approximating a superposition of delta peaks \cite{withTamas,withTadek}.
\par
When deriving the instability conditions from Eq. (\ref{EM}) in the following, we will focus on the particularly interesting case, where the mathematical expression under the root is non-negative (but the case of a negative value if $4M_{12}M_{21} < - (M_{11}-M_{22})^2$ could, of course, be treated as well). It can be shown that $\tilde{\lambda}$ becomes positive, if one of the following instability conditions is fulfilled:
\begin{equation}
M_{11} + M_{22} > 0 
\end{equation}
or 
\begin{equation}
 M_{11}M_{22} < M_{12}M_{21} \, .
\end{equation}
In this case, we expect the amplitudes of the small deviations $\delta p_i(x,t)$ to grow over time, which gives rise to spatial pattern formation (such as segregation). Inserting the abbreviations (\ref{abb1}) to (\ref{abb4}), the instability conditions become
\begin{equation}
\big[ A_{11} + (p_1^0P_{11}-D_1)\kappa^2 \big]
+ \big[ A_{22} + (p_2^0P_{22}-D_2) \kappa^2  \big] > 0 
\label{of1}
\end{equation}
and
\begin{eqnarray}
& & \big[ A_{11} + (p_1^0P_{11}-D_1)\kappa^2 \big]
 \big[ A_{22} + (p_2^0P_{22}-D_2)\kappa^2  \big] \nonumber \\
&<& \big( A_{12} + p_1^0P_{12}\kappa^2 \big) \big( A_{21} + p_2^0P_{21}\kappa^2 \big) \, . 
 \label{of}
\end{eqnarray}
If $\kappa$ is large enough (i.e. if the related cluster size is sufficiently small), the instability conditions (\ref{of1}) and (\ref{of}) simplify to
\begin{eqnarray}
p_1^0P_{11} + p_2^0P_{22} > D_1 + D_2 
\label{directed1}
\end{eqnarray}
and
\begin{equation}
(p_1^0P_{11}-D_1)(p_2^0P_{22}-D_2) < p_1^0P_{12} p_2^0P_{21}  \, .
\label{directed}
\end{equation}
These are further discussed in the main text.
\end{document}